# Tunable Doping and Mobility Enhancement in 2D Channel Field-Effect Transistors via Damage-Free Atomic Layer Deposition of AlOX Dielectrics

Ardeshir Esteki[1], Sarah Riazimehr[2], Agata Piacentini[1,3], Harm Knoops[2,4], Bart Macco[4], Martin Otto[3], Gordon Rinke[3], Zhenxing Wang[3], Ke Ran[3,5,6], Joachim Mayer[5,6], Annika Grundmann[7], Holger Kalisch[7], Michael Heuken[7,8], Andrei Vescan[7], Daniel Neumaier[3,9], Alwin Daus[1,10,*] and Max C. Lemme[1,3,*]

[1]Chair of Electronic Devices, RWTH Aachen University, Otto-Blumenthal-Str. 25, 52074 Aachen, Germany.

[2]Oxford Instruments Plasma Technology UK, North End, Yatton, Bristol, BS494AP, United Kingdom.

[3]AMO GmbH, Advanced Microelectronic Center Aachen, Otto-Blumenthal-Str. 25, 52074 Aachen, Germany.

[4] Department of Applied Physics, Eindhoven University of Technology, P.O. Box 513, 5600 MB Eindhoven, The Netherlands.

[5]Central Facility for Electron Microscopy GFE, RWTH Aachen University, Ahornstr. 55, 52074 Aachen, Germany.

[6]Ernst Ruska-Centre for Microscopy and Spectroscopy with Electrons (ER-C 2), Forschungszentrum Jülich, 52425, Jülich, Germany.

[7]Compound Semiconductor Technology, RWTH Aachen University, Sommerfeldstr. 18, 52074, Aachen, Germany.

[8]AIXTRON SE, Dornkaulstr. 2, 52134, Herzogenrath, Germany.

[9]Bergische Universität Wuppertal, Rainer-Gruenter-Str. 21, 42119 Wuppertal, Germany.

[10]Sensors Laboratory, Department of Microsystems Engineering, University of Freiburg, Georges-Köhler-Allee 103, 79110 Freiburg, Germany.

*Corresponding authors: alwin.daus@imtek.uni-freiburg.de; max.lemme@eld.rwth-aachen.de

## Abstract

Two-dimensional materials (2DMs) have been widely investigated because of their potential for heterogeneous integration with modern electronics. However, several major challenges remain, such as the deposition of high-quality dielectrics on 2DMs and the tuning of the 2DM doping levels. Here, we report a scalable plasma-enhanced atomic layer deposition (PEALD) process for direct deposition of a nonstoichiometric aluminum oxide (AlOX) dielectric, overcoming the damage issues associated with conventional methods. Furthermore, we control the thickness of the dielectric layer to systematically tune the doping level of 2DMs. The experimental results demonstrate successful deposition without detectable damage, as confirmed by Raman spectroscopy and electrical measurements. Our method enables tuning of the Dirac and threshold voltages of back-gated graphene and $MoS_2$ field-effect transistors (FETs), respectively, while also increasing the charge carrier mobility in both device types. We further demonstrate the method in top-gated $MoS_2$ FETs with double-stack dielectric layers ($AlOX+Al_2O_3$), achieving critical breakdown field strengths of 7 MV/cm and improved mobility compared with the back gate configuration. In summary, we present a PEALD process that offers a scalable and low-damage solution for dielectric deposition on 2DMs, opening new possibilities for precise tuning of device characteristics in heterogeneous electronic circuits.

**Introduction**

Two-dimensional materials (2DMs), such as graphene and transition metal dichalcogenides (TMDs), have great potential for heterogeneous integration with advanced silicon technology for future electronics [1–4]. Most prominently, the latter are considered as channel materials in ultimately scaled metal oxide semiconductor field-effect transistors (MOSFETs) [5–9]. Two key open challenges toward their application are the damage-free deposition of high-quality dielectrics with a high dielectric constant (high-κ) on the 2DMs and the controllable doping of the channel, which is crucial for controlling the threshold voltage of MOSFETs [10].

Plasma-enhanced atomic layer deposition (PEALD) is a production-ready technique for depositing high-κ dielectrics. Nevertheless, it often damages 2DMs, especially in the case of monolayers, due to highly reactive and energetic species formed during plasma activation [11,12]. The degree of damage incurred in the 2DMs can vary depending on the species used to create the plasma. While oxygen ($O_2$) is known to create highly reactive species that often damage 2DMs, especially graphene, other gases, such as hydrogen ($H_2$) or nitrogen ($N_2$), have been demonstrated to be less reactive. They induce a variable degree of degradation in 2DMs depending on the plasma power or the exposure time to the plasma [13–15]. This problem has been circumvented by using thin interlayers such as hexagonal boron nitride to protect graphene [12], with the drawback of complicating the fabrication by adding a second 2DM, hence requiring one additional transfer process. An alternative approach is to use PEALD to deposit hexagonal aluminum nitride on $WS_2$ as a seed layer [16], although some damage has still been reported throughout the process. Alternative methods to deposit uniform dielectrics onto 2DMs have been explored, including thermal ALD at low temperatures [17], seed layer deposition via physical vapor deposition followed by thermal ALD [18–22], the use of a growth substrate to enhance dielectric wetting during thermal ALD [23], and electron beam irradiation to carbonize the surface of graphene, which is then used as a seed layer during thermal ALD [24]. However, these approaches make depositing thin layers without increasing gate leakage difficult, which is critical for device performance.

Multiple methods for modifying the threshold voltage ($V_{Th}$) (as well as the Dirac voltage, $V_{Dirac}$, in the case of graphene) have been documented in the literature. For example, these voltages can be altered by modifying the substrate on which graphene is transferred [25,26] or by annealing 2DMs in phosphorous, sulfur, or vacuum environments [27–30]. Gallium irradiation has also been demonstrated to tune the $V_{Th}$ of MoS$_2$-channel FETs (MoS$_2$ FETs) [31]. However, after these modification methods are implemented, the 2D materials are still uncapped, i.e., they are exposed to atmospheric conditions, such as humidity. This exposure can unpredictably affect $V_{Th}$. Coating graphene with polymers, such as piperidine, has been shown to shift the $V_{Dirac}$, but this tuning effect diminishes quickly over time [32,33]. Electrostatic doping has been used to tune the $V_{Th}$ of MoS$_2$ [34], but its scalability is limited because of the need for a second gate. Self-assembled monolayers have also been shown to alter the $V_{Th}$ in MoS$_2$ devices, but in this case, a certain shift rather than controllable tuning has been demonstrated [35]. This means that the relationship between the treatment time and the tuning was not straightforward or linear, and consecutive shifts in a predictable pattern were not demonstrated. E-beam evaporated aluminum films oxidized in ambient air have also been used to control $V_{Th}$. However, this method is not easily scalable because it is not possible to precisely control oxidation under ambient conditions in a repeatable fashion [22]. Finally, e-beam evaporated dielectric layers have been employed to adjust the position of $V_{Th}$ [36]. Nevertheless, this approach introduces defects, leading to a reduction in the mobility of the 2DMs. In this work, we propose the tuning of $V_{Dirac}$ and $V_{Th}$ by depositing nonstoichiometric aluminum oxide films on top of 2DMs via a PEALD process. The method yields dielectrics with high critical breakdown fields and preserves the 2DM quality. We demonstrate our approach through detailed material analyses and electrical data from graphene- and molybdenum disulfide ($MoS_2$)-based devices. Our process is fully scalable to wafer dimensions, positioning PEALD as a promising and viable technique for the future introduction of 2DMs in semiconductor process lines.

**Results and discussion**

Our experiments are based on FETs and Hall bar structures with graphene and $MoS_2$ channels made by chemical vapor deposition. The fabrication process is illustrated in **Fig. 1a** and described in detail in the methods section. We use aluminum oxide deposited from a trimethylaluminum (TMA) precursor and an $O_2$ plasma, which has been extensively studied before [37–40] and will be labeled here as $Al_2O_3$ since it is close to that stoichiometry (see Supplementary Information (**SI) Fig. 1**). We also use a specifically developed aluminum oxide deposited using the same TMA precursor and a low-power nitrogen ($N_2$) plasma. Transmission electron microscopy (TEM) (**SI Fig. 2**) and X-ray photoelectron spectroscopy (XPS) (**SI Fig. 1**) analysis revealed that this resulted in a material containing N and carbon (C) atoms (see methods). For simplicity, we label this nonstoichiometric aluminum oxide (AlOX).

We fabricated two sets of graphene FETs (GFETs) with different dielectric configurations. The first set consisted of FETs with PEALD-deposited nonstoichiometric AlOX layers, with dielectric thicknesses of $t_{de} \approx$ 5 nm, 9 nm, and 13 nm. We fabricated a second set of FETs with an additional in situ PEALD-deposited $Al_2O_3$ layer to achieve a total gate dielectric thickness $t_{stack} \approx$ 18 nm. The $MoS_2$ FETs had gate dielectrics of $t_{de} \approx$ 4, 8, and 12 nm for AlOX and total thicknesses of $t_{stack} \approx$ 14 nm, 17 nm, and 19 nm, respectively. A schematic of a bottom-gate device with a 2DM channel is shown in **Fig. 1b**. This configuration was experimentally investigated with graphene and $MoS_2$ as the 2DMs. **Fig. 1c** displays an optical image of a $MoS_2$ FET.

The image in **Fig. 2a** shows a cross-sectional transmission electron microscopy (TEM) image of a Si-SiO$_2$-Graphene-AlOX transistor with an AlOX thickness of 13 nm. The effect of dielectric deposition on 2DMs was investigated via Raman spectroscopy (**Fig. 2b**). Statistical analyses of the D-to-G peak intensity ratios and the full width-half-maxima of the 2D peak indicate that no measurable process damage occurs in graphene. In contrast, the standard Al$_2$O$_3$ plasma deposition process performed as a control yields a significantly increased D peak and larger FWHM (**Fig. 2c, Fig. 2d**). Details of the XPS, TEM and Raman experiments are described in the Methods section. We observed a doping effect of the graphene channel in back-gate voltage-dependent drain current measurements ($I_D$ vs. $V_{BG}$) (**Fig. 2e**), which were performed under ambient conditions in a four-point probe (4p) configuration according to **Fig. 1b** and **Fig. 1c**. The measurements of 14 devices revealed strong p-doping of graphene before deposition, with $V_{Dirac} \approx 30$ V. After dielectric deposition, $V_{Dirac}$ shifted in the negative direction by ~ 25, ~ 35 and ~ 44 V for $t_{de}$ values of 5, 9 and 13 nm, respectively (**Fig. 2f**). The maximum electron and hole mobilities were extracted via the direct transconductance method [41], which excludes contact resistance effects due to the 4p configuration (see Methods section for details) [42]. In the case of GFETs, the dielectric capping layer increases the maximum hole mobility from ~ 1500 to ~ 2300 cm$^2$V$^{-1}$s$^{-1}$ and the maximum electron mobility from ~ 150 to ~ 1600 cm$^2$V$^{-1}$s$^{-1}$ (**Fig. 2g**).

We carried out a similar experiment for back-gated Si-SiO$_2$-MoS$_2$-AlOX transistors. The Raman spectra confirmed negligible damage to this semiconducting 2D material as well (**Fig. 3a**). We also observed a doping effect for MoS$_2$, as shown in the $I_D$ vs. $V_{BG}$ measurements in the 4p configuration in **Fig. 3b**. Measurements in air performed on 7 MoS$_2$ FETs show $V_{th} \approx 1$ V before deposition and a shift in the negative direction to $V_{th}$ values of ~ -3, ~ -7 and ~ -11 V for thicknesses of $t_{de}$ = 4, 8 and 12 nm, respectively (**Fig. 3c**). In this case, the mobility, extracted at a constant overdrive voltage of 37.5 V, increased from ~ 0.2 $cm^2V^{-1}s^{-1}$ to ~ 4 $cm^2V^{-1}s^{-1}$ (**Fig. 3d**).

The observed negative shifts in $V_{Dirac}$ and $V_{th}$ with AlOX deposition point to the presence of positive fixed charges in the dielectric. The fixed charges in ALD $Al_2O_3$ have been extensively studied, especially within the context of surface passivation layers [43]. The amount of charge intricately depends on the preparation method and substrate. On silicon, $Al_2O_3$ routinely yields negative fixed charges [44]. Specifically, PEALD $Al_2O_3$ prepared with $O_2$ plasma has strong negative fixed charges on the order of $-5 \times 10^{12}$ $cm^{-2}$, whereas thermal ALD $Al_2O_3$ prepared with $H_2O$ has fewer negative fixed charges on the order of $-10^{11}$ $cm^{-2}$ [39]. These negative charges have been reported to be located at the interface between Si and $Al_2O_3$ [38]. Conversely, PEALD $Al_2O_3$ on germanium has no appreciable charge [45]. In our case, the $N_2$ plasma approach likely resulted in positive bulk charges. While the exact origin of these positive fixed charges is subject to further study, it could be related to the presence of C and N in the film, which are negligible when $O_2$ plasma is employed. Importantly, the use of $N_2$ plasma not only enables damage-free deposition on 2DMs but also plays a strong role in the tunability of $V_{Dirac}$ and $V_{th}$ [46–49].

The increase in mobility can be attributed to a reduction in phonon and Coulomb scattering and a reduction in water and oxygen molecules adsorbed on the surface of the 2DMs [50–53]. On the other hand, it has also been shown that the strong increase in current and mobility can be due to the presence of positive fixed charges in the AlOX layer. The positive fixed charges in the dielectric layer cause an excess of electrons to accumulate at the interface between the graphene or $MoS_2$ channel and the AlOX, which in turn causes an increase in the on-current and mobility of the 2DMs [22,48,49].

We then proceeded to fabricate top-gated devices using the AlOX+Al$_2$O$_3$ double stack as a dielectric layer (**Fig. 4a**, **SI Fig. 3a**). The Al$_2$O$_3$ layer was added to increase the dielectric strength, i.e., the breakdown electrical field, and to reduce the gate leakage currents. This was necessary, as devices employing only AlOX showed significant leakage currents in the top-gate configuration. The dielectric constant of the gate stack was extracted via the double-gate current–voltage method, which yielded a dielectric constant of ~ 14 for the MoS$_2$ FETs (**SI Fig. 3b–d**; for details, see Methods section) [54]. Top-gated 4p measurements ($I_D$-$V_{TG}$) (**SI Fig. 4a**) yielded a hysteresis of ~0.7 V in the MoS$_2$ FETs (**SI Fig. 4b**). The graph in **SI Fig. 4c** shows the progression of the top gate leakage current with increasing top gate voltage. The leakage current causes an irreversible breakdown at 7.5 V, corresponding to a breakdown electric field strength of ~ 6.5 MV/cm, which aligns with values reported for similar AlOX-Al$_2$O$_3$ stacks in the literature [17,19,55–57]. It is also comparable to those obtained for high-quality dielectrics deposited by ALD directly on silicon substrates [37], which indicates the high quality of the films with a low amount of pinholes. We performed 4p measurements for 5 devices each with gate stack thicknesses of t$_{stack}$ ~ 14, ~ 17, and ~ 19 nm and extracted threshold voltages of V$_{th}$ ~ -2, ~ -6, and ~ -8 V (**Fig. 4b, Fig. 4c**) and mobilities of ~ 10, ~ 17 and ~ 23 cm$^2$V$^{-1}$s$^{-1}$ (**Fig. 4d**), respectively. Details about the thickness and dielectric composition ratios can be found in the Methods section. These results show that threshold voltage tuning is also possible in top-gated devices and that the beneficial effect on mobility is maintained even as the devices are controlled with the top gate. This is an important extension of the method, as the PEALD process enables top-gate and gate-all-around/surround-gate MoS$_2$ FETs, which are under investigation for the ultimate extension of Moore's law.

A comparison between our work and the state-of-the-art methods for tuning the doping of 2DMs in FETs is shown in **Table 1**. While several methods have been demonstrated to influence $V_{Dirac}$ or $V_{Th}$, either by shifting them or by tuning them, our approach offers a unique combination of benefits. Our PEALD process is a scalable technique that, as demonstrated by Raman spectroscopy, does not damage 2DMs; it enhances their mobility while providing encapsulation. Furthermore, the precise thickness control of ALD allows controlling the doping in both graphene and MoS$_2$, alongside their $V_{Dirac}$ and $V_{Th}$, respectively.

**Conclusions**

We demonstrated a scalable and low-damage PEALD process for directly depositing high-quality dielectric films on 2DMs. By precisely controlling the thickness of the AlOX layer, we control the amount of positive fixed charges in the dielectrics, which enables tuning of the positions of the $V_{Dirac}$ and $V_{th}$ of the graphene and MoS$_2$ field-effect transistors. Furthermore, the process was used in combination with stoichiometric Al$_2$O$_3$ to fabricate top-gated graphene and MoS$_2$ FETs with high breakdown voltages and increased mobilities compared with the respective back-gated devices. This work opens new pathways for the heterogeneous integration of devices with 2DMs, preserving material quality and controlling their doping.

**Methods**

**Device fabrication**

Highly p-doped silicon (Si) chips with dimensions of 2x2 cm$^2$ and 90 nm silicon dioxide (SiO$_2$) grown by thermal oxidation were used as substrates. The bottom contacts were patterned via negative optical lithography using a double layer of resist (LOR 3A, micro resist technology) and AZ5214E photoresist (MicroChemicals). For the graphene devices, 5 nm of titanium and 15 nm of palladium were deposited via electron-beam evaporation, followed by lift-off in dimethyl sulfoxide (DMSO). For the molybdenum disulfide (MoS$_2$) devices, 20 nm of nickel was deposited via electron-beam evaporation.

Commercially available CVD-grown single-layer graphene on copper (Cu) (Grolltex, San Diego, USA) and multilayer (~ 4–7 layers) MoS$_2$ materials grown in-house by metal–organic vapor–phase epitaxy (MOVPE) on sapphire were used in this work.

Graphene field-effect transistors (GFETs) and MoS$_2$ FETs were fabricated by transferring either graphene or MoS$_2$ from the growth substrate to the patterned bottom contacts via a poly(methyl methacrylate) (PMMA) frame-assisted wet transfer process. In the case of graphene, the copper was etched in an HCl-H$_2$O$_2$-H$_2$O solution, and the frame was then rinsed in DI water before being transferred to the patterned samples. In the case of MoS$_2$, the material was delaminated from the growth substrate in DI water via mechanical stress.

A positive lithography step using an AZ5214E photoresist was performed to pattern the graphene. Low-power reactive ion etching (RIE) with oxygen plasma was used to etch the graphene. The resist was subsequently stripped in acetone at 50°C for 60 minutes.

MoS$_2$ was patterned via a positive lithography step with a double stack of LOR 3A and AZ5214E on top of the PMMA from the transfer. After development, the MoS$_2$ remained covered by PMMA in the etched region. The PMMA was then etched by low-power RIE with oxygen plasma, after which the MoS$_2$ was etched in a fluorine and oxygen plasma. The LOR 3A and AZ5214E resists were stripped in a TMAH-containing developer after UV flood exposure [58]. Finally, the underlying PMMA layer was removed through annealing at 400°C under vacuum.

After transfer, AlOX layers were deposited on the samples via trimethylaluminum (TMA) as a precursor and N$_2$ plasma. The process was performed in an Oxford Instruments Atomfab PEALD system with a remote RF-driven capacitively coupled plasma (CCP) plasma source and a grounded table [59]. In this system, plasma generation electrodes and counter electrodes are placed remotely from the substrate surface, and the grounded substrate is not part of the plasma generation area [patent application PCT/GB2019/052763] [60]. The deposition was carried out at a table temperature of 200°C and a source RF forward power of 60 W. The chamber gas mixture used a ratio of 3:4 for plasma N$_2$ gas to plasma Ar gas. The growth cycle time was approximately 1.3 s, with each cycle consisting of an N$_2$/Ar plasma time of 100 ms. The layers had thicknesses of 5, 9, and 13 nm for the back-gated GFETs and 4, 8 and 12 nm for the back-gated MoS$_2$ FETs.

A second set of devices was prepared for the top-gated FETs. For the graphene FETs, a nonstoichiometric AlOX layer of 8 nm was complemented with a stoichiometric Al$_2$O$_3$ layer of 10 nm (total thickness $t_{stack}$~18 nm). For the MoS$_2$ FETs, an AlOX layer of 7 nm and an Al$_2$O$_3$ layer of 5 nm were used (total thickness $t_{stack}$~12 nm). A third set of devices was prepared for the top-gated MoS$_2$ FET doping experiments. In this case, AlOX layers with thicknesses of 4, 7 and 9 nm were employed, while the Al$_2$O$_3$ thickness was fixed at 10 nm, resulting in total thicknesses $t_{stack}$ ~ 14, 17, and 19 nm, respectively. The Al$_2$O$_3$ layer was deposited in the same Oxford Instruments Atomfab ALD system using TMA and oxygen plasma as precursors. The deposition temperature was 300°C, and the power was set to 100 W.

The top gates were patterned via negative optical lithography using an AZ5214E photoresist. In the case of the graphene devices, 5 nm of titanium and 15 nm of palladium were deposited, whereas 20 nm of palladium were deposited via electron-beam evaporation in the case of the $MoS_2$ FETs.

**Material Characterization**

### Transmission Electron Microscopy

Cross-sectional transmission electron microscopy (TEM), energy-filtered TEM (EFTEM) and energy dispersive X-ray spectroscopy (EDX) were conducted with a Tecnai F20 and a JEOL-JEM200 at an operating voltage of 200 kV. Lamellas for this analysis were cut from the respective GFETs with a Strata400 focused ion beam system using gallium.

The image in **SI Fig. 2a** shows a TEM cross-section of a fabricated graphene device. By energy-dispersive X-ray spectroscopy (EDX), we identified a layer containing Al, measuring approximately 10 nm in thickness (**SI Fig. 2b**). EFTEM allowed us to determine the distributions of Carbon (C), nitrogen (N) and oxygen (O), which had comparable thicknesses (**SI Fig. 2c-e**). The elevated O levels of the film imply that, despite the use of $N_2$ plasma for depositing the dielectric layer, the resulting layer is likely composed of AlOX. The presence of O can be attributed to the slow deposition rate of the dielectric, which presumably leads to the incorporation of O from background gases and water ($H_2O$) into the thin film.

### Raman Spectroscopy

Raman spectroscopy was conducted via a WiTec alpha300R Raman imaging microscope equipped with a laser at a wavelength of 532 nm at a laser power of 1 mW. The examination was performed before and after the deposition of the dielectric layers to monitor the quality of the 2DMs. For graphene, measurements were performed on $7 \times 40$ µm$^2$ maps, each containing 280 spectra, at various locations for each sample. The resulting spectra were subjected to Lorentzian function fitting, allowing extraction of the ratio between the D and G peak intensities and the full width at half maximum (FWHM) of the 2D peak. For MoS$_2$, single-point measurements were performed.

### X-Ray Photoelectron Spectroscopy

X-Ray photoelectron spectroscopy (XPS) was performed with a Thermo Scientific KA1066 spectrometer and monochromatic Al K-α X-rays at 1486.6 eV. Charging effects in the XPS spectra were considered when applying charge correction to position the C−C bonding contribution from adventitious carbon at 284.8 eV. Contributions from O1s, Al2p, N1s and Si2p regions were measured. Furthermore, an elemental depth profile was obtained by Ar+ ion sputtering, which confirmed the nonstoichiometric composition of the deposited AlOX dielectric (~30% Al, ~59% O, ~7% C, ~4% N) (**SI Fig. 1**).

## Electrical Characterization

### Hall Bars

Hall bars with six terminals were employed as test structures, whereas the devices functioned as three-terminal field-effect transistors. The inner terminals (depicted in **Fig. 1b** and **Fig. 1c**) were used to measure the effective voltage difference $V_{\text{diff}}$, accounting for the contact resistance, which was then employed to estimate the mobility of the devices.

The electrical measurements of the FETs were carried out via a Cascade Summit 12000 semiautomatic probe station connected to a Hewlett Packard 4156B precision semiconductor parameter analyzer and a Hewlett Packard E5250A low leakage switch mainframe. The carrier mobility for both global back-gated devices and top-gated devices was calculated via four-terminal measurements via the following equation:

$$\mu = (\partial(I_{\text{DS}}/V_{\text{diff}})/\partial V_{\text{BG}}) \cdot L_{12} d_{\text{dielectric}}/(W_{channel} \cdot \varepsilon_{\text{r}} \cdot \varepsilon_0)$$

where $I_{DS}$ is the drain-source current across $V_{diff}$ is the voltage difference between the inner contacts, $V_{GS}$ is the voltage difference between the gate and source, $d_{dielectric}$ is the thickness of the gate dielectric, $L_{12}$ is the distance between the inner contacts, W is the width of the channel, κ is the dielectric constant of the gate dielectric and $\varepsilon_0$ is the vacuum permittivity. The channel measures approximately 42 μm in length and 10 μm in width (11 μm in width in the case of MoS$_2$), with a drain-source voltage of 100 mV.

**Double-Gate Measurements**

We employed a dual-gate configuration to extract the dielectric constant (**SI Fig. 3a**), utilizing both a top gate and a back gate to modulate the graphene channel [54]. The plot in **SI Fig. 3b** shows the top gate transfer curves for different back-gate voltages. The top-gate was swept between -5 V and +5 V, whereas the back-gate was held constant at 0 V, +10 V, -10 V, +20 V, -20 V, +30 V, and -30 V for each transfer curve. The top-gate voltage at the charge neutrality point was plotted as a function of the back-gate voltage, and the shift was then fitted to determine the ratio of the top-gate capacitance ($C_{TG}$) to the back-gate capacitance ($C_{BG}$) (**SI Fig. 3c**).

The following equation was used to calculate the dielectric constant (κ) of the AlOX-Al$_2$O$_3$ stack:

$$slope = \frac{C_{BG}}{C_{TG}}$$

where $C_{BG}$ is the capacitance of the back-gate and $C_{TG}$ is the capacitance of the top-gate. This equation can be simplified to:

$$\kappa_{TG} = {t_{TG} \cdot \kappa_{BG}}/{(t_{BG} \cdot slope)}$$

where $\kappa_{TG}$ and $\kappa_{BG}$ = 3.9 are the dielectric constants of the top and back gate dielectrics, respectively, while $t_{TG}$ = 12 nm and $t_{BG}$ = 90 nm are their respective thicknesses (**SI Fig. 3d**).

**Acknowledgments**

This work has received funding from the European Union's Horizon 2020 research and innovation programme under grant agreement No 952792 (2D-EPL) and No 881603 (Graphene Flagship Core 3), by the German Federal Ministry of Education and Research (BMBF) within the project 03XP0210 (GIMMIK) and by the German Research Foundation through the Emmy Noether programme (506140715).

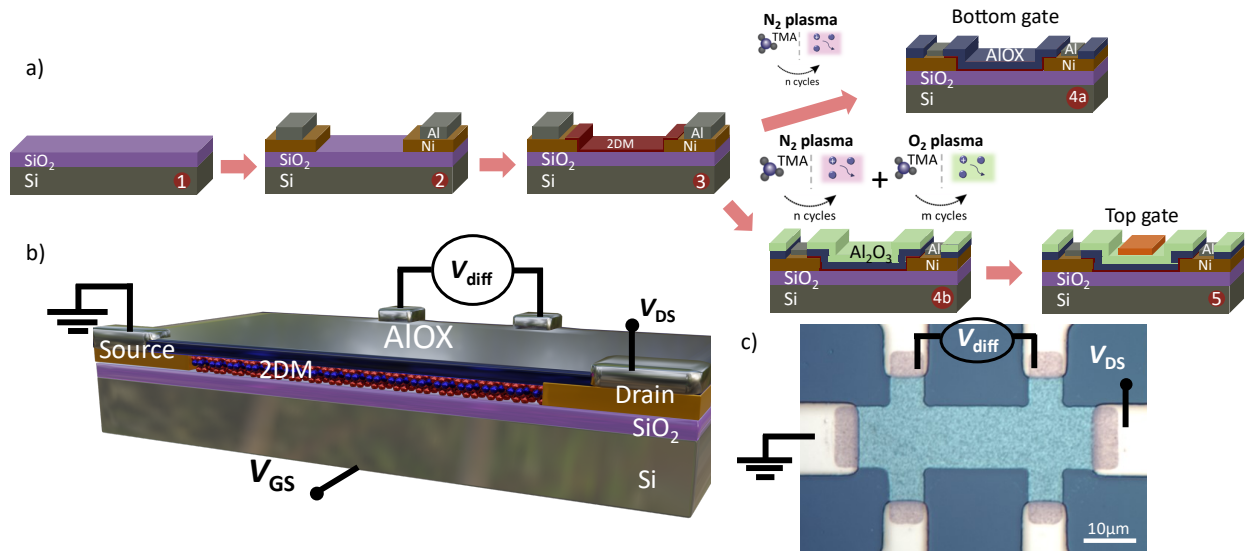

**Figure 1: 2D Transistor Fabrication Processes and Device Structure**

**a)** Process used to fabricate the FETs (see methods). FETs in two configurations have been fabricated: bottom gate FETs with an AlOX top dielectric encapsulation (4a) and top gate FETs with an AlOX+$Al_2O_3$ gate dielectric stack (4b, 5). **b)** Cross-sectional schematic representation of a back-gated FET in a four-point probe configuration (see methods), incorporating a Ni/Al source and drain bottom contacts and an AlOX dielectric layer. **c)** Top-view optical image of the back-gated device. The channel measures approximately 42 µm in length and 11 µm in width.

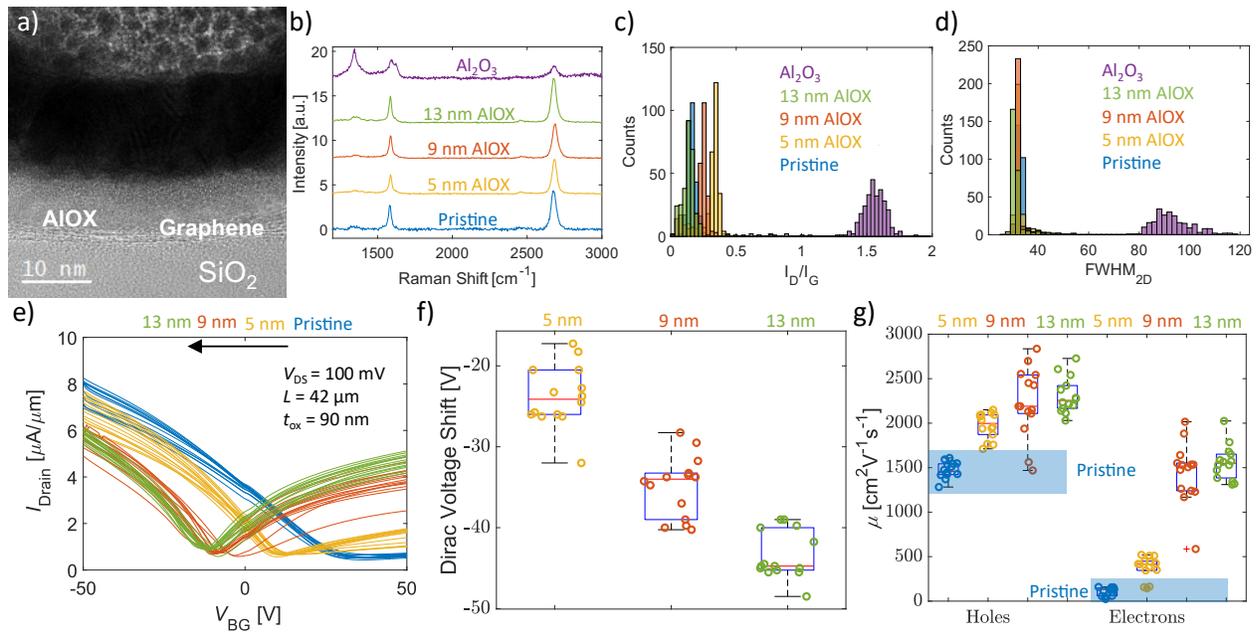

**Figure 2: Graphene Back-Gated Field-Effect Transistors with AlOX Dielectric**

**a)** Cross-sectional TEM image of a Si-SiO$_2$-Graphene-AlOX transistor with an AlOX thickness of 13 nm. **b-d)** Raman analysis of graphene under various conditions, including measurements in air and after the deposition of AlOX layers at $t_{de}$ values of 5 nm, 9 nm, and 13 nm, in comparison to those of graphene after the deposition of Al$_2$O$_3$. The Raman maps include data collected from 210 points. **b)** Single Raman spectra. **c)** Ratio of the D peak intensity to the G peak intensity. **d)** Full width at half maximum (FWHM) of the 2D peak. **e-g)** Comparison of back-gated electrical data using a four-point probe measurement configuration (see methods), both before and after the deposition of AlOX layers at $t_{de}$ values of 5 nm, 9 nm, and 13 nm. The channel measures approximately 42 μm in length and 10 μm in width, with a drain-source voltage of 100 mV: **e)** Current–voltage transfer curves. **f)** Measured shift in the Dirac voltage due to deposition of the AlOX layer. **g)** Evaluation of hole and electron mobilities.

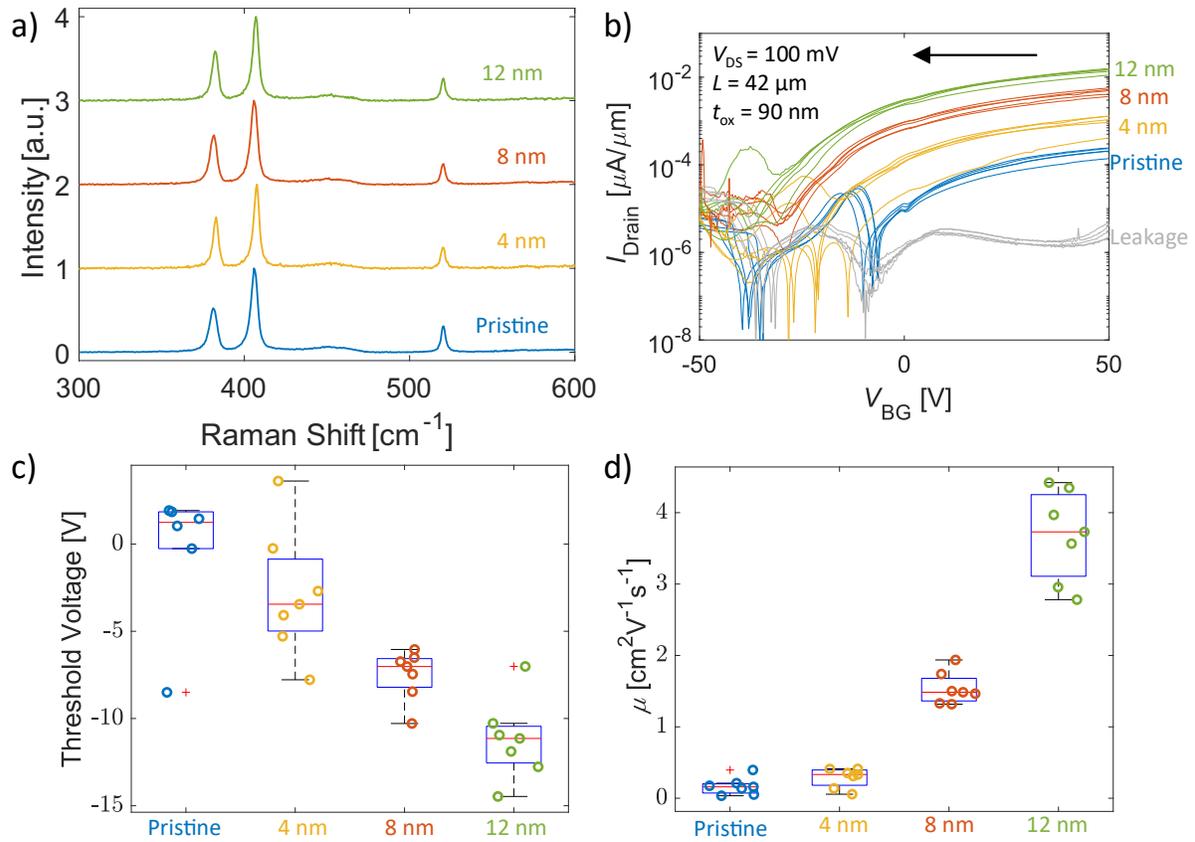

**Figure 3: Back-Gated MoS$_2$ Field-Effect Transistors**

**a)** Raman spectra of MoS$_2$, comparing measurements taken in air and after AlOX deposition at t$_{de}$ values of 4 nm, 8 nm, and 12 nm. **b-d)** Comparative analysis of back-gated electrical data before and after AlOX deposition at t$_{de}$ values of 4, 8, and 12 nm. The channel measures approximately 42 μm in length and 11 μm in width, with a drain-source voltage of 100 mV: **b)** Current–voltage transfer curves (semilogarithmic scale). **c)** Threshold voltage position. **d)** Mobility assessment measured at an overdrive voltage of 37.5 V for each device.

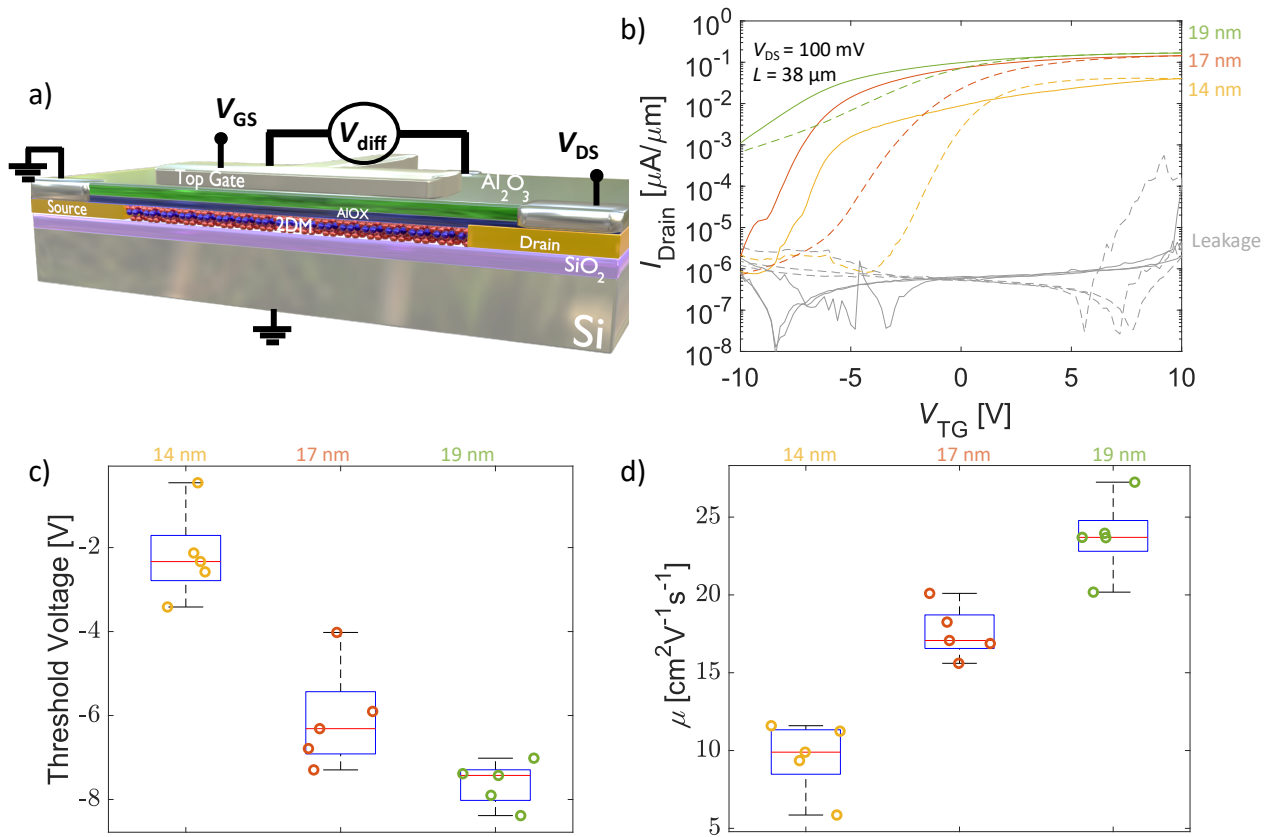

**Figure 4: Top-Gated MoS$_2$ Field-Effect Transistors**

Comparison of top-gated electrical data obtained via a four-point probe measurement configuration (see methods) for three sets of devices with AlOX+Al$_2$O$_3$ dielectric layers at t$_{stack}$ values of 14, 17 and 19 nm. **a)** Cross-sectional schematic representation of a top-gated MoS$_2$ FET. The channel measures approximately 42 μm in length and 11 μm in width, with a drain-source voltage of 100 mV: **b)** Current–voltage transfer curves (semilogarithmic scale). Continuous lines correspond to forward sweeps, whereas dashed lines correspond to backward sweeps. The leakage current is shown in gray. **c)** Threshold voltage position. **d)** Mobility assessment measured at an overdrive voltage of 6 V for each device.

| Method | Channel material | Type of doping | Mode | Capping | Scalable | Mobility | Reference |
|---|---|---|---|---|---|---|---|
| FIB Gallium Irradiation | $MoS_2$ | **Tuning** | Back Gate | No | No | Decreases | 31 |
| Sulfur/Hydrogen Treatment | $MoS_2$ | Shift | Back Gate | No | **Yes** | **Increases** | 29 |
| Vacuum/Air Annealing | $MoS_2$ | Shift | Back Gate | No | **Yes** | Decreases | 30 |
| Electrostatic Double Gate | $MoS_2$ | **Tuning** | Back Gate | **Yes** | No | Unchanged | 34 |
| E-beam Evaporation | $MoS_2$ | Shift | Back Gate | **Yes** | **Yes** | Decreases | 36 |
| SAM | $MoS_2$ | Shift | Back Gate | **Yes** | **Yes** | Decreases | 35 |
| E-beam Evaporation | $MoS_2$ | Shift | Back Gate | **Yes** | No | **Increases** | 22 |
| Substrate Trapping | Graphene | Shift | Back Gate | No | **Yes** | Unchanged | 26 |
| Bottom Gate Dielectric Composition Modulation | Graphene | **Tuning** | Back Gate | No | **Yes** | - | 25 |
| Molecular Charge Trapping | Graphene | **Tuning** | Back Gate | **Yes** | **Yes** | Decreases | 33 |
| Ammonia Annealing | Graphene | Shift | Back Gate | No | **Yes** | Decreases | 27 |
| Phosphorous/Nitrogen Annealing | Double Layer Graphene | **Tuning** | Back Gate | No | **Yes** | **Increases** | 28 |
| **PEALD AlOX** | **Graphene/$MoS_2$** | **Tuning** | Back Gate | **Yes** | **Yes** | **Improves** | *This Work* |
| | | | Top Gate | **Yes** | **Yes** | **Increases** | |

**Table 1.** Comparison of the state-of-the-art methods for tuning the Dirac voltage in graphene FETs and the threshold voltage in $MoS_2$ FETs. Studies on $MoS_2$ are highlighted with a gray background, graphene studies with a blue background, and double-layer graphene studies with an orange background. Our work meets all the criteria and is highlighted in green.

# Tunable Doping and Mobility Enhancement in 2D Channel Field-Effect Transistors via Damage-Free Atomic Layer Deposition of AlOX Dielectrics

Ardeshir Esteki[1], Sarah Riazimehr[2], Agata Piacentini[1,3], Harm Knoops[2,4], Bart Macco[4], Martin Otto[3], Gordon Rinke[3], Zhenxing Wang[3], Ke Ran[3,5,6], Joachim Mayer[5,6], Annika Grundmann[7], Holger Kalisch[7], Michael Heuken[7,8], Andrei Vescan[7], Daniel Neumaier[3,9], Alwin Daus[1,10,*] and Max C. Lemme[1,3,*]

[1]Chair of Electronic Devices, RWTH Aachen University, Otto-Blumenthal-Str. 25, 52074 Aachen, Germany.

[2]Oxford Instruments Plasma Technology UK, North End, Yatton, Bristol, BS494AP, United Kingdom.

[3]AMO GmbH, Advanced Microelectronic Center Aachen, Otto-Blumenthal-Str. 25, 52074 Aachen, Germany.

[4] Department of Applied Physics, Eindhoven University of Technology, P.O. Box 513, 5600 MB Eindhoven, The Netherlands.

[5]Central Facility for Electron Microscopy GFE, RWTH Aachen University, Ahornstr. 55, 52074 Aachen, Germany.

[6]Ernst Ruska-Centre for Microscopy and Spectroscopy with Electrons (ER-C 2), Forschungszentrum Jülich, 52425, Jülich, Germany.

[7]Compound Semiconductor Technology, RWTH Aachen University, Sommerfeldstr. 18, 52074, Aachen, Germany.

[8]AIXTRON SE, Dornkaulstr. 2, 52134, Herzogenrath, Germany.

[9]Bergische Universität Wuppertal, Rainer-Gruenter-Str. 21, 42119 Wuppertal, Germany.

[10]Sensors Laboratory, Department of Microsystems Engineering, University of Freiburg, Georges-Köhler-Allee 103, 79110 Freiburg, Germany.

*Corresponding authors: alwin.daus@imtek.uni-freiburg.de; max.lemme@eld.rwth-aachen.de

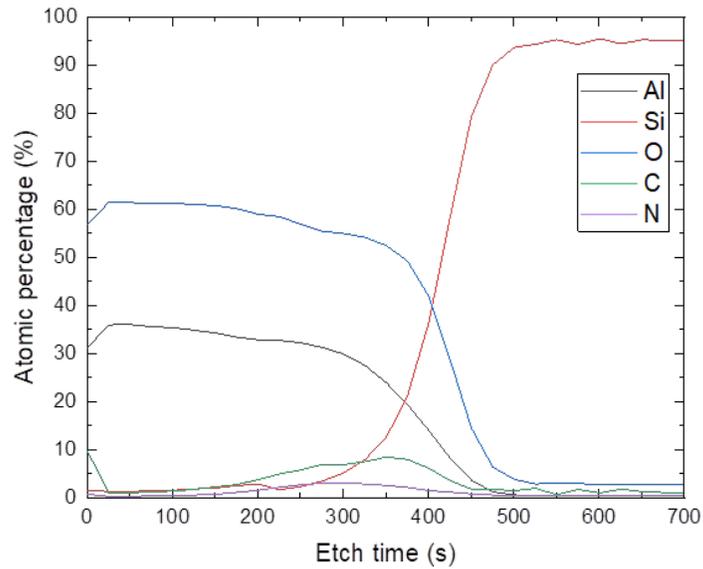

**SI Figure 1: XPS depth profiling**

X-ray photoelectron spectroscopy depth profile of a Si-AlOX-$Al_2O_3$ dielectric stack with an AlOX thickness of ~ 9 nm and an $Al_2O_3$ thickness of ~ 10 nm. Disregarding the initial data point, which is influenced by adventitious carbon (C) contamination from the environment, the subsequent measurements reveal an oxygen (O) content near 60% and an aluminum (Al) content of approximately 36%, which indicates $Al_2O_3$ stoichiometry for the top dielectric layer. The AlOX layer shows increased levels of C and N with respect to the top $Al_2O_3$ layer. Note that the exact profile can be affected by preferential ion sputtering effects, as well as convolution, since the XPS depth sensitivity is on the same order of magnitude as the $Al_2O_3$ and AlOX layer thicknesses.

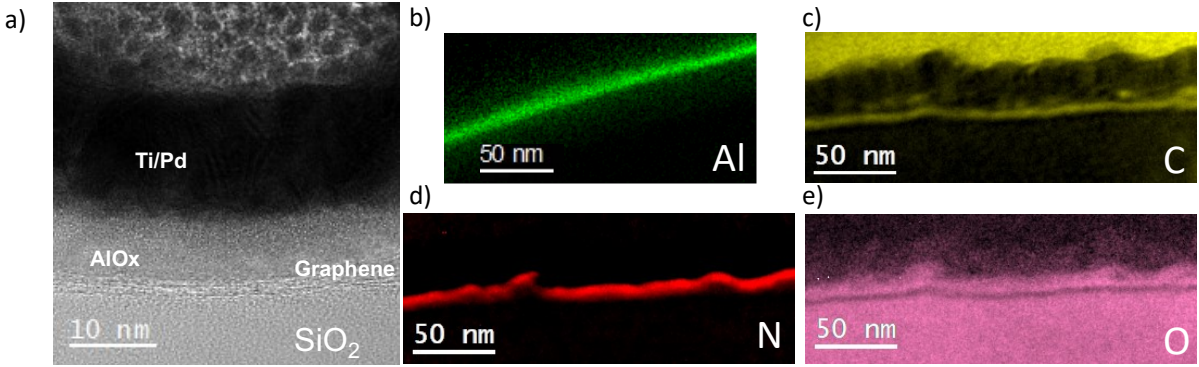

**SI Figure 2: AlOX material analysis**

**a-e)** Structural and elemental analysis by transmission electron microscopy (TEM) taken at three different positions. **a)** Cross-sectional TEM image of a Si-SiO$_2$-Graphene-AlOX transistor with an AlOX thickness of 13 nm. **b)** EDXs – energy dispersive X-ray spectroscopy showing the distribution of aluminum (~ 13 nm). **c-e)** Energy-filtered TEM image from the same region showing the distributions of **c)** carbon (~ 3.7 nm), **d)** nitrogen (~ 10 nm) and **e)** oxygen atoms (~ 10 nm above C).

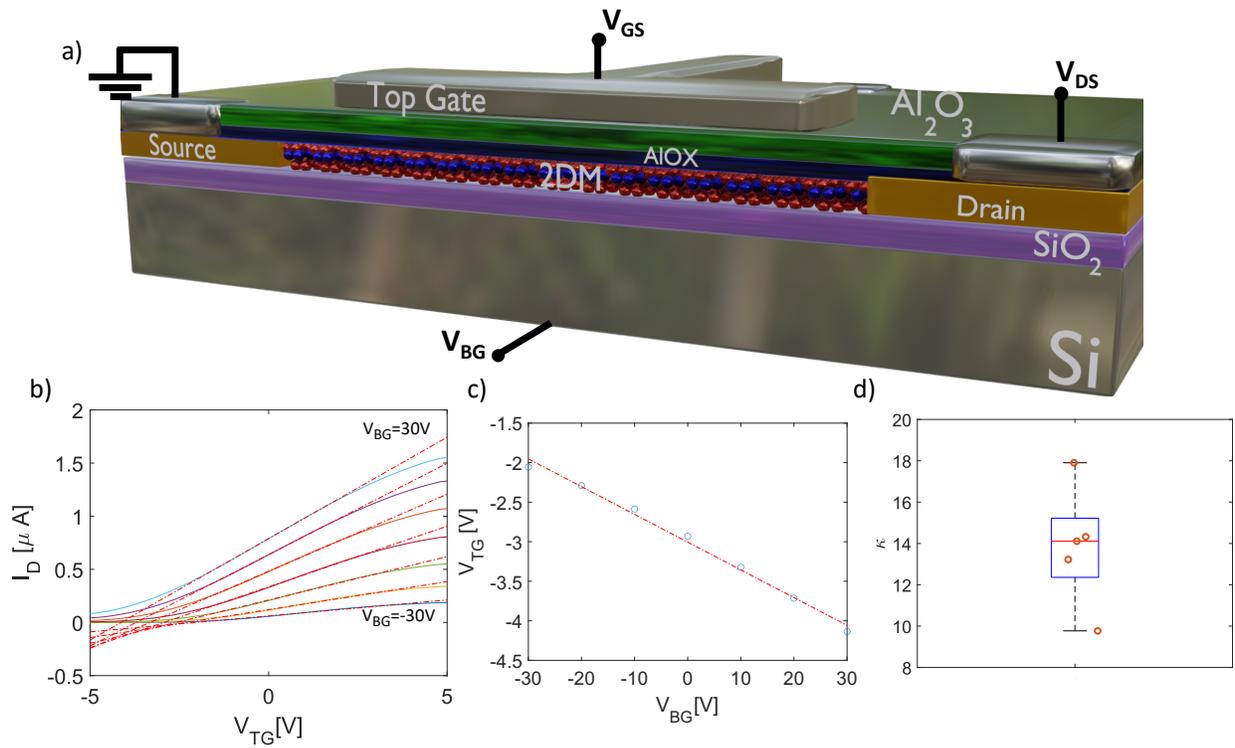

**SI Figure 3. AlOX-Al$_2$O$_3$ Dielectric Constant Extraction**

**a)** Schematic representation of a back-gated MoS$_2$ FET in a two-point configuration, incorporating source and drain bottom contacts and an AlOX-Al$_2$O$_3$ dielectric stack. **b)** Top-gated current–voltage curves (sweep: -5 V to +5 V) while keeping the back-gate voltage constant. Each curve represents a different back-gate voltage from -30 V to +30 V with steps of 10 V. **c)** Top gate threshold voltage as a function of the applied back-gate voltage. **d)** Statistical extraction of the dielectric constant via the double gate measurement (see methods) [54].

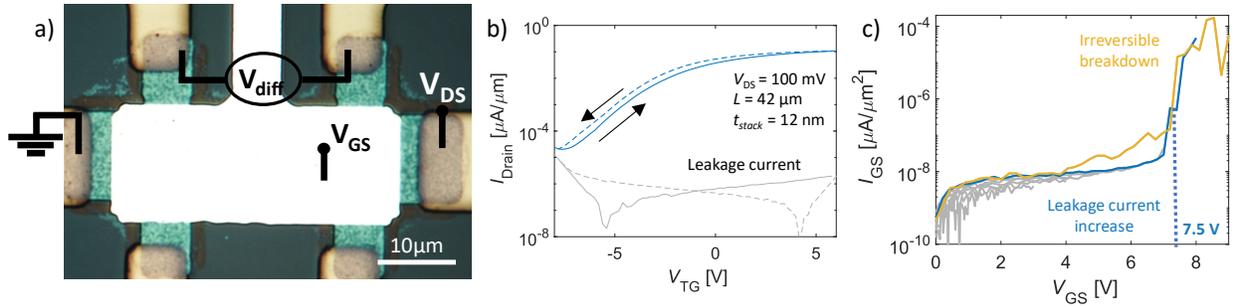

**SI Figure 4: Top-Gated MoS$_2$ Field-Effect Transistors**

Top-gated electrical data from MoS$_2$ FETs in the four-point probe configuration. **a)** Top-view optical image. **b)** Top-gated current–voltage transfer curve. The top gate was swept between -7 V and + 6 V with a step size of 200 mV. The continuous line corresponds to forward sweeps, whereas the dashed lines correspond to backward sweeps. The leakage current is shown in gray. **c)** Top-gate leakage current progression with increasing applied gate voltage for one device. The blue line shows a significant increase in leakage current, whereas the yellow line shows irreversible breakdown at 7.5 V. The gray lines show the measurements performed before the dielectric breakdown.